
\input phyzzx
\def\rarrow{\rightarrow}
\def\({[}
\def\){]}

\REF\wein{S. Weinberg, Phys. Rev. Lett. 37 (1976) 657.}
\REF\glwe{S.L. Glashow and S. Weinberg, Phys. Rev. D15 (1977) 1958.}
\REF\dogo{J.F. Donoghue and E. Golowich, Phys. Rev. D37 (1988) 2542.}
\REF\dln{C.O. Dib, D. London and Y. Nir,
 Int. J. Mod. Phys. A6 (1991) 1253.}
\REF\bsg{E. Thorndike, CLEO Collaboration, a talk given in
the meeting of the American Physical Society, Washington D.C. (1993).}
\REF\hhg{For a detailed review of multi-scalar models and
a guide to the literature, see: J.F. Gunion, H.E. Haber,
G.L. Kane and S. Dawson, {\it The Higgs Hunter's Guide}
(Addison-Wesley Publishing Company, Reading, MA, 1990).}
\REF\bbg{G.C. Branco, A.J. Buras and J.-M. Gerard, Nucl. Phys. B259
(1985) 306.}
\REF\fran{For a review of $B-\bar B$ mixing and a guide to the
literature, see: P.J. Franzini, Phys. Rep. 173 (1989) 1.}
\REF\krpo{P. Krawczyk and S. Pokorski, Nucl. Phys. B364 (1991) 10.}
\REF\niqu{For a review of
 CP asymmetries in $B$ decays and a guide to the literature, see:
 Y. Nir and H.R. Quinn, Ann. Rev. Nucl. Part. Sci.
 42 (1992) 211.}
\REF\nisi{Y. Nir and D. Silverman, Nucl. Phys. B345 (1990) 301.}
\REF\pdg{K. Hikasa et al. (Particle Data Group), Phys. Rev. D45
(1992) S1.}
\REF\bhp{V. Barger, J.L. Hewett and R.J.N. Phillips,
 Phys. Rev. D41 (1990) 3421.}
\REF\buras{A.J. Buras, P. Krawczyk, M.E. Lautenbacher and C. Salazar,
 Nucl. Phys. B337 (1990) 284.}
\REF\krpob{P. Krawczyk and S. Pokorski, Phys. Rev. Lett. 60 (1988) 182;
 \hfill\break H. Isidori, Phys. Lett. B298 (1993) 409.}
\REF\nircpv{Y. Nir, Lectures presented in the 20th SLAC Summer Institute,
 SLAC-PUB-5874 (1992).}
\REF\gsw{B. Grinstein, R. Springer and M. Wise,
 Nucl. Phys. B339 (1990) 269.}
\REF\thdbsg{W.-S. Hou and R.S. Willey, Nucl. Phys. B326 (1989) 54;
 \hfill\break T.G. Rizzo, Phys. Rev. D38 (1988) 820.}
\REF\hewett{J.L. Hewett, Phys. Rev. Lett. 70 (1993) 1045; \hfill\break
 V. Barger, M.S. Berger and R.J.N. Phillips,
Phys. Rev. Lett. 70 (1993) 1368.}
\REF\bisa{I.I. Bigi and A.I. Sanda, Phys. Rev. Lett. 58 (1987) 1604;
\hfill\break H.-Y. Cheng, Phys. Rev. D42 (1990) 2329.}
\REF\chengt{H.-Y. Cheng, Int. J. Mod. Phys. A7 (1992) 1059.}
\REF\nirsarid{Y. Nir and U. Sarid, Phys. Rev. D47 (1993) 2818.}
\REF\jarl{C. Jarlskog, Phys. Rev. Lett. 55 (1985) 1039.}

\def\Weizmann{\centerline{\it Weizmann Institute of Science}
\centerline{\it Physics Department, Rehovot 76100, Israel}}

{\baselineskip=11pt
\Pubnum={WIS-93/51/Jun-PH}
\date={June, 1993}
\titlepage
\vskip 2cm
\title{{\bf Implications of $b\rightarrow s\gamma$ for CP Violation
 in Charged Scalar Exchange}}
\vskip 2cm
\author{Yuval Grossman and Yosef Nir
\foot{Incumbent of the Ruth E. Recu Career Development Chair
supported in part by the Israel Commission for Basic Research,
and by the Minerva Foundation.}}
\Weizmann
\vskip 2cm

\centerline{\bf Abstract}
In models of three or more scalar doublets, new CP violating phases
appear in charged scalar exchange. These phases affect CP asymmetries
in neutral $B$ decays, even if Natural Flavor Conservation holds.
The recent upper bound on the decay $b\rightarrow
s\gamma$ constrains the effect to be at most of order a few percent.
Modifications of constraints on the CKM parameters open an interesting
new region in the $\sin2\alpha-\sin2\beta$ plane even in the absence
of new phases.
\vfill
\centerline{Submitted to {\sl Physics Letters B}}
}
\endpage

A detailed investigation of $B$-meson decays is a promising way
to discover or severely constrain New Physics beyond the
Standard Model. In particular, $B$ physics is sensitive to
extensions of the scalar sector: in most models
scalars couple more strongly to heavier quarks and thus may
affect bottom (and top) decays while having a negligible effect
on lighter quark decays.

Models of three or more scalar doublets allow for new CP violating
phases in charged scalar exchange $\(\wein\)$. It is conventional
wisdom that if Natural Flavor Conservation (NFC) $\(\glwe\)$
is imposed on the Yukawa couplings, then these phases do not affect
CP asymmetries in neutral $B$ decays $\(\dogo,\dln\)$.
We show that this is not the case: the new phases have an effect on
$B-\bar B$ mixing and consequently on CP asymmetries. All existing
bounds from CP violating processes -- $\epsilon$, $\epsilon^\prime$
and $D_n$ (the electric dipole moment of the neutron) -- do not exclude
strong effects in CP asymmetries in $B$ decays. However, the
CP violating couplings contribute also to the (CP conserving) radiative
decay $B\rightarrow X_s\gamma$. We find that the recent upper bound
on this decay $\(\bsg\)$ does constrain the shift in
CP asymmetries to be small, at most 0.02.

We also investigate the modifications
in the predictions for the CP asymmetries that result from the
different constraints on CKM parameters. These arise because there
are contributions from scalar mediated diagrams to
$B-\bar B$ mixing and to $\epsilon$. We find that
a combined measurement of the CP asymmetries in $B$ decays
into $\psi K_S$ and $\pi\pi$ may probe new contributions to
$B-\bar B$ mixing even if there are no new phases involved.

To explain how the new effects arise and to study how they are
constrained by experimental data, it is simplest to work in
the framework of three doublet models, and to assume that
one of the two charged scalars is much heavier than the other.
However, our results hold for any multi-scalar model (with
at least three doublets) where
NFC is implemented by requiring that only one doublet
couples to each quark sector and that it is a different one
in each sector (model II $\(\hhg\)$).
The couplings of the two physical
charged scalars $H_i^\pm$ ($i=1,2$) to quarks are given by
(see {\it e.g.} ref. $\(\bbg\)$)
$${\cal L}_H={G_F^{1/2}\over 2^{1/4}}\sum_{i=1,2}
\left\(H_i^+\bar U(Y_i M_u V (1-\gamma_5)+X_i V M_d(1+\gamma_5))D
+{\rm h.c.}\right\).\eqn\hcouplings$$
Here $V$ is the CKM matrix, while $X$ and $Y$ are complex numbers that
depend on mixing
parameters in the charged scalar sector. The contributions to
$B-\bar B$ mixing from box diagrams with intermediate $W$-bosons
and $H_1$-scalars (we assume that the heavier scalar $H_2$
contributes negligibly) are of the form $\(\fran\)$
$$M_{12}^B={G_F^2\over64\pi^2}(V_{td}^* V_{tb})^2\(I_{WW}
+I_{HH}+2I_{WH}\).\eqn\monetwo$$
The Standard Model contribution is $I_{WW}$ while box diagrams with
two $H_1$-propagators or one $H_1$- and one $W$-propagator give
$I_{HH}$ and $I_{HW}$, respectively. The potentially
large and interesting contributions are contained in $I_{HH}$:
$$\eqalign{
I_{WW}=&\ m_t^2 I_0(x_t)V_{LL},\cr
I_{HH}\approx&\ m_t^2y_t\(|Y|^4 I_1(y_t) V_{LL}+
(XY^*)^2 y_b I_2(y_t) S_{LL}\),\cr}\eqn\iwwhh$$
where $x_q\equiv m_q^2/m_W^2$, $y_q\equiv m_q^2/m_H^2$, and $m_H$
is the mass of the lightest charged scalar.
The $I_i$ functions are given by
$$\eqalign{
I_0(x)=&\ 1+{9\over 1-x}-{6\over(1-x)^2}-{6x^2\ln x\over(1-x)^3},\cr
I_1(y)=&\ {1+y\over(1-y)^2}+{2y\ln y\over(1-y)^3},\cr
I_2(y)=&\ {8\over(1-y)^2}+{4(1+y)\ln y\over(1-y)^3}.\cr}\eqn\ifunctions$$
We calculate the various matrix elements in the vacuum insertion
approximation $\(\krpo\)$:
$$\eqalign{
V_{LL}\equiv&\ \bra{\bar B}(\bar d\gamma^\mu(1-\gamma_5)b)^2
\ket{B}={4\over3}f_B^2 m_B,\cr
S_{LL}\equiv&\ \bra{\bar B}(\bar d(1-\gamma_5)b)^2
\ket{B}\approx-{5\over6}f_B^2 m_B.\cr}\eqn\via$$
In eq. \iwwhh\ we gave only the terms most relevant to our argument.
However, in our calculations we use the full expressions.

Within the Standard Model, CP asymmetries in $B$ and $B_s$
decays depend on the angles $\alpha$, $\beta$ and $\gamma$
of the unitarity triangle only $\(\niqu\)$. In the multi-scalar
framework, they are affected by the new phase in charged
scalar exchange through the dependence of $B-\bar B$ mixing on
Im$(XY^*)$. (The effects on decay amplitudes are negligible.)
The argument that there is no such effect results
from approximating $y_b=0$ in eq. \iwwhh. This is not always justified.
What $I_{HH}$ (and $I_{WH}$) really depend on is the {\it Yukawa
coupling} to the bottom quark, ${m_b/\VEV{\phi_d}}$. It is a rather
attractive option in multi scalar models to have ${m_b\over m_t}
\sim{\VEV{\phi_d}\over\VEV{\phi_u}}$. In such a case, the Yukawa
coupling of the bottom quark is as large as that of the top quark,
and the term proportional to $y_b$ is as important as the one
that is not. It is in this region of parameter space
that CP violation from charged scalar exchange may
have large effects on CP asymmetries in $B$ decays.
Note that in the neutral $K$ system, the corresponding terms are
proportional to the Yukawa coupling of the strange quark and therefore
the CP violating effect is negligible.

{}From eq. \iwwhh\  we can learn in what
region of the $(X,Y)$ parameter space the new effects may
play a significant role. If $|X|\lsim|Y|$, then the term proportional to
$I_1$ dominates $I_{HH}$. This terms contributes with the same phase as
$I_{WW}$ and thus the Standard Model predictions for the CP asymmetries
are not modified. A necessary condition for large effects is then
$${|X|\over|Y|}\gsim{m_t\over m_b}.\eqn\conxy$$
However, if \conxy\ is fulfilled with $|XY|\lsim1$, then $I_{WW}$
dominates over $I_{HH}$ and the Standard Model predictions are
still unchanged. Thus we also need
$$|XY|\gg1.\eqn\conxyb$$

To present the way in which the Standard Model predictions are modified,
we define a phase
$$\theta_H=\arg(I_{WW}+I_{HH}+2I_{HW}).\eqn\defth$$
Then, if the Standard Model predicts that a certain asymmetry
equals sin($\theta_{SM}$), in a multi scalar model the
prediction is modified to sin($\theta_{SM}+\theta_H$), {\it e.g.}
$$\eqalign{
a_{CP}(B\rarrow \psi K_S)=&\ -\sin(-2\beta+\theta_H),\cr
a_{CP}(B\rarrow \pi^+\pi^-)=&\ \sin(2\alpha+\theta_H).\cr}\eqn\modify$$
(The overall minus sign in $a_{CP}(B\rarrow \psi K_S)$
arises because the final $\psi K_S$ state is CP-odd.)
If the phase $\theta_H\ll\theta_{SM}$,
the shift $\Delta a_{CP}$ from the Standard Model prediction is
$$\Delta a_{CP}\lsim\sin(\theta_H).\eqn\delacp$$
A few points are in order:
\par a. The CKM phase is factored out in \monetwo,
so that $I_{WW}$ is real. Then, when $I_{HH}$ and $I_{HW}$ are
real, the Standard Model result is reproduced, as it should.
\par b. As the modification is in the phase of the
mixing amplitude and not in the decay amplitude, the shift
of the measured angle is universal, namely independent of
the decay mode.
\par c. Eqs. \conxy\ and \conxyb\
imply that the effect is never very large
in $B_s$ decays: When $|X|$ is large, an additional term in $I_{HH}$,
$$I_{HH}^q=\ m_t^2y_by_q|X|^4I_1(y_t)V_{LL},\eqn\qdepend$$
($q=d$ or $s$ for $B_d$ or $B_s$, respectively) becomes important in
$B_s$ decays, and it contributes with the same phase as the
Standard Model diagram. For very large $|X|$,
the corrections in $B_s$ decays are small. Consequently,
the angles ``$\beta$", ``$\alpha$" and $\gamma$ deduced naively
from $B\rarrow\psi K_S$, $B\rarrow\pi\pi$ and $B_s\rarrow\rho K_S$,
respectively, will sum up to approximately $\pi$, even though the
first two do not correspond to angles of the unitary triangle
anymore. This is an example of a general result that holds
when the phase of $B_s-\bar B_s$ mixing is the same as in the
Standard Model $\(\nisi\)$.

To see if indeed large effects in $B$ decays are possible, we now study
the experimental constraints on $X$ and $Y$.
For the charged scalar mass, we use $m_H\gsim m_Z/2$ $\(\pdg\)$.
Both $|X|$ and $|Y|$ are
constrained by the requirement of perturbativity $\(\bhp\)$:
$$|X|\lsim120,\ \ |Y|\lsim6,\ \ \Longrightarrow\ \
{\rm Im}(XY^*)\leq|XY|\lsim720.\eqn\pert$$
The value of $|Y|$ is constrained by $B-\bar B$ mixing $\(\bhp,\buras\)$,
$$|Y|\lsim\cases{2&$m_H\sim m_Z/2$,\cr 3&$m_H\sim 2m_Z$.\cr}\eqn\bbound$$
The value of $|X|$ is constrained by $B\rarrow X\tau\nu$ $\(\krpob\)$,
$$|X|\lsim{m_H\over 0.54\ GeV},\eqn\taubound$$
but only if it is one and the same doublet scalar that couples to
the charged leptons and to down quarks. We thus consider below
also values of $|X|$ that do not fulfill \taubound.
A direct bound on Im($XY^*$) comes from CP violating processes.
The strongest among these comes from $D_n$ $\(\krpo,\nircpv\)$,
$${\rm Im}(XY^*)\lsim\cases{
20&$m_H\sim m_Z/2$,\cr 100&$m_H\sim 2m_Z$.\cr}\eqn\dnbound$$
(This bound arises from quark electric dipole moment operators.
Bounds from the three gluon operator may be stronger, but
suffer from larger hadronic uncertainties.)

The strongest constraint on Im($XY^*)$, however, comes -- somewhat
surprisingly\foot{This situation was actually foreseen in ref.
$\(\krpo\)$.}  -- from a CP conserving process, the
decay $b\rarrow s \gamma$ $\(\bsg\)$:
$$BR(B\rarrow X_s\gamma)\leq5.4\times10^{-4}.\eqn\bsgexpt$$
Within multi-scalar models with NFC, this branching ratio is given by
$\(\krpo\)$
$$BR(B\rarrow X_s\gamma)=C|\eta_2+G_W(x_t)+(|Y|^2/3)G_W(y_t)
+(XY^*)G_H(y_t)|^2,\eqn\bsgtheory$$
where
$$C\equiv{3\alpha\eta_1^2\ BR(B\rightarrow X_c\ell\nu)\over
2\pi F_{ps}(m_c^2/m_b^2)}\approx 3\times10^{-4}.\eqn\cvalue$$
$F_{ps}\sim0.5$ is a phase space factor, $\eta_1\sim0.66$ and
$\eta_2\sim 0.57$ are QCD corrections factors $\(\gsw\)$.
The expressions for the $G$-functions can be found in ref. $\(\krpo\)$.
In the two Higgs-doublet model $\(\gsw,\thdbsg\)$,
$XY^*=1$ and the bound \bsgexpt\ gives a lower bound on
$m_H$ almost independently of $Y$ $\(\hewett\)$. In the three
Higgs-doublet model, $m_H\sim m_Z/2$ is still allowed.

\FIG\figA{The upper bound on Im($XY^*)$ as a function of the lightest
charged scalar mass $m_H$. The three curves correspond to
$m_t=90$ (solid), 140 (dashed) and 180 (dotted) $GeV$.}

%

The upper bound on Im($XY^*)$ corresponds to a case where the
real part of the new diagrams cancels the Standard Model contributions
and the upper bound \bsgexpt\ is saturated by the imaginary part
of these diagrams:
$${\rm Im}(XY^*)\lsim\sqrt{5.4\times10^{-4}\over C}\ {1\over G_H(y_t)}.
\eqn\upper$$
 The results are presented in Fig. \figA. For $m_t\sim140\ GeV$ we get
$${\rm Im}(XY^*)\lsim\cases{
2&$m_H\sim m_Z/2$,\cr 4&$m_H\sim 2m_Z$.\cr}\eqn\bsgbound$$
For heavier (lighter) top mass, the bounds are stronger (weaker).

The upper bound on Im$(XY^*)$ implies that charged scalar exchange
can make only a negligible contribution to $\epsilon$ and
$\epsilon^\prime$ and cannot be the {\it only} source of CP violation.
(For recent attempts to allow this possibility see ref. $\(\bisa\)$.
The combination of the upper bound on $D_n$ and the lower
bound on $m_H$ made it very unlikely $\(\krpo,\nircpv,\chengt\)$.)
On the other hand, the contribution to $D_n$ can still be close
to the experimental upper bound.

\FIG\figB{The upper bound on $\theta_H$, the shift in the phase measured
in CP asymmetries in $B$ decays (see \defth), as a function of
arg($XY^*)$ for $m_H\sim m_Z/2$ and $Y=0.1$. The three curves correspond
to $m_t=90$ (solid), 140 (dashed) and 180 (dotted) $GeV$.}

%

The bound \bsgbound\ by itself does not imply that the effects
of charged scalar exchange on CP asymmetries are small. However,
the upper bound on $b\rightarrow s\gamma$ also implies that the
conditions \conxy\ and \conxyb\ cannot be simultaneously fulfilled.
Take, for example, the case that $|Y|\ll1$.
Then the term proportional to $|Y|^2$ in \bsgtheory\ is negligible.
An upper bound on $|XY^*|$ is derived when arg$(XY^*)\sim\pi$:
$|XY^*|\lsim2$, in contradiction to \conxyb.
A survey of the $(X,Y)$ values consistent with \bsgexpt\ leads to the
results shown in Fig. \figB. We find
$$\theta_H\leq1.2^o\ \Longrightarrow\  \Delta a_{CP}\lsim0.02.
\eqn\maxeffect$$
The effect is smaller for heavier charged scalar or for $Y$-values
different from those presented in Fig. 2. An effect of the magnitude
\maxeffect\ is still larger than the hadronic uncertainties
in $B\rightarrow\psi K_S$. However, it is
too small to be unambiguously observed in
$B$-factories, where the accuracy in $a_{CP}(B\rightarrow\psi K_S)$
is expected to be of ${\cal O}(0.05)$ $\(\niqu\)$.

\FIG\figC{The allowed region in the $\sin2\alpha-\sin2\beta$ plane
in the Standard Model (solid) and in multi-scalar models (dot-dashed).}

%

Modifications of the Standard Model predictions
for CP asymmetries in $B$ decays may also arise from the different
constraints on CKM parameters. This holds even for two scalar
doublet (type I and type II) models where indeed there are no new phases.
The most significant effect is that the lower bounds
on $|V_{tb}V_{td}^*|$ from $B-\bar B$ mixing and from $\epsilon$
are relaxed, because charged scalar exchange may contribute
significantly $\(\buras\)$. This opens up a region in the plane of
$\sin 2\alpha-\sin2\beta$ forbidden in the Standard Model,
as shown in Fig. \figC. We used here the same input parameters
as in ref. $\(\nirsarid\)$.\foot{We improved upon the analysis
of ref. $\(\nirsarid\)$ by working in the three dimensional parameter
space of $(\rho,\eta,|V_{cb}|)$ rather than integrating over the allowed
range for $|V_{cb}|$. We find that the Standard Model
lower bound on $\sin2\beta$ is even stronger than
the one given in $\(\nirsarid\)$: $\sin2\beta\geq0.23$.}
We find an interesting result, which goes
beyond the specific extension of the Standard Model investigated here:
If experiment finds a relatively low
value of $\sin2\beta$ (below 0.5) and a {\it negative} value of
$\sin2\alpha$, it may be an indication that there are significant
contributions from new physics to $B-\bar B$ mixing, even if these
contributions carry no new phases! We should also emphasize
that a nice feature of the Standard Model -- that at low $\sin2\beta$
values there is a strong correlation between $\sin2\beta$ and
$\sin2\alpha$ and, in particular, $|\sin2\alpha|$ cannot be small --
is maintained even in the presence of the new effects discussed here.

In terms of the CP-violating measure $J$ $\(\jarl\)$,
the multi scalar-doublet model allows lower values:
$$J\geq\cases{1.4\times10^{-5}&Standard Model,\cr
5.5\times10^{-6}&Multi-Scalar,\cr}\eqn\newj$$
while the upper bound, $J\leq6.3\times10^{-5}$, remains unchanged.

To summarize, in models of three or more scalar doublets and natural
flavor conservation, new
phases in charged scalars exchange may affect CP asymmetries in
$B$ decays. The recent upper bound on the decay
$B\rarrow X_s\gamma$ constrains both CP violating and CP conserving
charged scalar couplings, implying that the
deviations from the Standard Model predictions cannot exceed a
few percent.

\refout
\endpage
\figout

\end